\documentclass[a4paper,10pt]{article}
\usepackage[utf8x]{inputenc}

\title{Non-anticommutativity in Presence of a Boundary  }
\author{Mir Faizal $^1$ and Douglas J Smith $^2$\\
$^1$ Mathematical Institute, University of Oxford, 
 \\ Oxford,
OX1 3LB, United Kingdom. \\
$^2$ Department of Mathematical Sciences, Durham University, 
\\ Durham, DH1 3LE, United Kingdom.}

\begin{document}

\maketitle

\begin{abstract}
In this paper we consider non-anticommutative field theories in $\mathcal{N} =2$ superspace 
formalism on three-dimensional manifolds with a boundary. We  modify the original Lagrangian    
in such a way that it preserves half the supersymmetry 
even in the presence of a boundary. We also analyse the partial  breaking  
of  supersymmetry caused by  non-anticommutativity between 
 fermionic coordinates. 
Unlike in four dimensions, in three dimensions a theory with $\mathcal{N} =1/2$ supersymmetry 
cannot be obtained by a non-anticommutative deformation of an $\mathcal{N} =1$
theory.
However, in this paper we construct a
three dimensional 
theory with $\mathcal{N} =1/2$ supersymmetry by
studying a combination of non-anticommutativity and boundary effects,
starting from $\mathcal{N} =2$ supersymmetry.
\end{abstract}

\section{Introduction}
There are many interesting deformations of field theories which can be
realized
on the worldvolume of D-branes in various string theory backgrounds.
The presence of a constant $NS-NS$ B-field background gives rise to noncommutativity 
 \cite{a, b, c, Chu:1998qz,Chu:1999gi, d}.
The concept of noncommutative coordinates can be extended to superspace
\cite{Ferrara:2000mm, Klemm:2001yu, Buchbinder:2001at, Samsonov:2001an}.
This concept of spacetime noncommutativity can be extended to include more
general deformations of the (super-)Poincar\'e algebra
\cite{Kosinski:2000xu, Britto:2003ey, Ivanov:2003te}, and for Grassmann
coordinates this leads to non-anticommutativity \cite{Ferrara:2000mm, Klemm:2001yu, e, Britto:2003aj, f, Britto:2003kg, Lunin:2003bm, g, Ferrara:2004zv, h}.
Such deformations are realised on the worldvolume of D-branes in
$RR$ backgrounds \cite{Ooguri:2003tt, 2, 1,3,4}, and the
gravity dual of such a field theory has been constructed in \cite{Chu:2008qa}.
Also, a graviphoton background gives rise to a noncommutativity between
spacetime and superspace coordinates \cite{2, 1a, 2a, 3a, 4a}.
Noncommutative deformations generated by the $NS-NS$ and graviphoton backgrounds do not break any 
supersymmetry. However, the non-anticommutative deformation breaks the supersymmetry corresponding to the
deformed superspace coordinate. In four dimensions
it is possible to break the supersymmetry generated by one Weyl supercharge while leaving
the supersymmetry generated by the other intact. Thus, if we start from a theory 
with $\mathcal{N} =1$ supersymmetry
in four dimensions and perform a non-anticommutative deformation, 
we can arrive at a theory with  $\mathcal{N} =1/2$
supersymmetry. Now, a theory with $\mathcal{N} =1$ supersymmetry in four dimensions has the same amount of
supersymmetry as a theory with $\mathcal{N} =2$ supersymmetry in three dimensions.
 Thus, from a three dimensional perspective 
this corresponds to breaking the supersymmetry from  $\mathcal{N} =(1,1)$ to
$\mathcal{N} =(0,1)$ or $\mathcal{N} =(1,0)$
supersymmetry \cite{4abcde}. Furthermore, a theory with $\mathcal{N} =1$ supersymmetry in three 
dimensions has the same amount of supersymmetry as a
theory with $\mathcal{N} =2$ supersymmetry in two dimensions. However, we cannot carry this argument further, as there are not 
enough degrees of freedom to perform this non-anticommutative deformation in
two dimensions without breaking all supersymmetry. So, we cannot partially break 
supersymmetry to $\mathcal{N} =1/2$ supersymmetry in three dimensions by non-anticommutative deformations alone. 
However, we will show in this paper that we can obtain a theory with $\mathcal{N} =1/2$ supersymmetry in three dimensions 
by combining the non-anticommutative deformations with boundary effects. 

In determining the Euler-Lagrange equations of a Lagrangian field theory one encounters
 terms
which can be written as a surface integral. In theories that are at most quadratic in derivatives
 this is the only contribution that remains when an action is varied and its Euler-
Lagrange equations are used. Thus, in the presence of a boundary one must specify boundary conditions 
that ensure the above surface term vanishes.
The boundary
breaks translation invariance and so it also breaks  supersymmetry. 
In fact, the supersymmetric transformation of most theories transforms into a surface 
term and this generates a boundary term in the presence of a boundary. 
This problem can be eliminated by imposing boundary conditions under which this boundary term vanishes. 
However, the bulk theory can also be modified by introducing a boundary action such that its supersymmetry 
transformations exactly cancel the boundary term generated by the supersymmetry transformations of the original
bulk action. This way half of the original supersymmetry is 
preserved.  This has been done for three dimensional theories in $\mathcal{N} =1$ 
superspace \cite{eb, eb1111}. 
Such boundary effects for M2-branes have also been analysed in  $\mathcal{N} =1$
 superspace \cite{fb,gb, hb}.
In this paper we will first generalize these results to a three dimensional 
theory in $\mathcal{N} =2$ 
superspace and then 
 analyse the non-anticommutative deformation of this theory. 
We will thus be able to arrive at a theory with $\mathcal{N} =1/2$ supersymmetry 
in three dimensions.
 As non-anticommutativity
occurs due to the coupling of D-branes to  $RR$ fields, it would be interesting to study 
 a  non-Abelian Born-Infeld  Lagrangian in this
 non-anticommutative superspace.
In this context the boundary 
effects analysed in this paper could be used to study a system of
D2-branes ending on D4-branes in the presence of  $RR$ fields. 
With this motivation we consider the example of
 a flat-space Born-Infeld Lagrangian \cite{1b, 2b, 3b, 4b} coupled to scalar matter in three-dimensional $\mathcal{N} =2$ superspace. See \cite{Gates:1983nr}
for a useful review of three-dimensional superspace.

\section{Boundary Supersymmetry }
In this section we review the method of introducing a boundary action in order
to preserve half the supersymmetry without explicit boundary conditions. We
also define our notation. This
was originally carried out for $\mathcal{N} = 1$ in \cite{eb}
and extended to $\mathcal{N} = 2$ in \cite{fb}.

We start from an $\mathcal{N} =1$ superfield  $\phi(\theta)$, where $\theta$ is a two component Grassmann parameter. It  
 transforms under supersymmetric transformations as  
$$
 \delta \phi(\theta) =  \epsilon^a Q_a\phi(\theta),
\;\; \rm{where} \;\;
 Q_a = \partial_a  - (\gamma^\mu \theta)_a \partial_\mu,
$$
is the generator of $\mathcal{N} =1$ supersymmetry. 
If in component form  the field has the following form 
$$
\phi(\theta) = p + q \theta + r \theta^2,
$$
 then the supersymmetric 
transformation can be written as
\begin{eqnarray}
  \delta p &=&\epsilon^a q_a, \nonumber \\
  \delta q_a &=& -\epsilon_a r  + (\gamma^\mu\epsilon)_a \partial_a p, \nonumber \\
  \delta r &=& \epsilon^a (\gamma^\mu \partial_\mu)_a^b q_b.
\end{eqnarray}
Now the Lagrangian for an $\mathcal{N} =1$ theory can be written in terms of
such a superfield as 
\begin{equation}
 \mathcal{L} = D^2 [\phi(\theta)]_{\theta =0},
\end{equation}
where $D^2 = D^a D_a/2$ and $ D_{a} = \partial_{a} + (\gamma^\mu \theta)_a \partial_\mu $. 
This Lagrangian is invariant under these supersymmetric transformations on a manifold without boundaries. However, 
if there is a boundary, say at $x_3 =0$, then the supersymmetric transformations of
the Lagrangian are given by 
$
 \delta \mathcal{L} = - \partial_3 ( \epsilon \gamma^3 q)
$.
This breaks the supersymmetry of the resultant theory. 
However, if we add or subtract the following term 
$
 \mathcal{L}_b =  \partial_3 [\phi(\theta)]_{\theta =0}
$, 
to the original Lagrangian, then the supersymmetric transformation of the total Lagrangian is given by 
\begin{equation}
 \delta [ \mathcal{L} \pm  \mathcal{L}_b] = \pm 2  \partial_3 \epsilon^{\pm} q_{\mp},
\end{equation}
where $q_{\pm} = P^{\pm}q \equiv ( 1 \pm \gamma^3)  q/2 $.  Hence, we can preserve the supersymmetry generated by either   
$\epsilon^- Q_+$ or $\epsilon^+ Q_-$ by adding or subtracting $\mathcal{L}_b$ to $\mathcal{L}$. However, 
we cannot preserve all the supersymmetry. Thus, the Lagrangian  which preserves the supersymmetry corresponding to 
$\epsilon^- Q_+$ is $\mathcal{L}^+$, and to $\epsilon^+ Q_-$ is $\mathcal{L}^-$
where
\begin{equation}
\mathcal{L}^\pm =\mathcal{L} \pm \mathcal{L}_b =  (D^2 \mp \partial_3) [\phi]_{\theta =0}.
\end{equation}

After reviewing  boundary supersymmetric  theories in $\mathcal{N} =1$
superspace formalism, we present the straightforward generalisation of these
results to 
theories with $\mathcal{N} =2$ supersymmetry. 
Thus,  we will analyse a Lagrangian with $\mathcal{N} =2$ supersymmetry, 
\begin{equation}
 \mathcal{L} = D^2_1 D^2_2  [\Phi (\theta_1, \theta_2)]_{\theta_1 =\theta_2 =0},
\end{equation}
where  $ D_{1a} = \partial_{1a} + (\gamma^\mu \theta_1)_a \partial_\mu,$ and $ D_{2a} = \partial_{2a} + (\gamma^\mu \theta_2)_a \partial_\mu,$
 are the standard covariant derivatives which commute with $Q_{1a}$ and $Q_{2a}$,
and $\Phi$ is an $\mathcal{N} =2$ scalar superfield.
We can decompose a superfield with $\mathcal{N} =2$ supersymmetry, into two copies 
of $\mathcal{N} =1$ superfields.  So, we can write $\Phi (\theta_1, \theta_2)$ as 
\begin{eqnarray}
 \Phi (\theta_1, \theta_2)&=& p_1 ( \theta_1) + q_1 (\theta_1) \theta_2 + r_1 (\theta_1) \theta_2^2
\nonumber \\ 
 &=& p_2 ( \theta_2) + q_2 (\theta_2) \theta_1 + r_2 (\theta_2) \theta_1^2,
\end{eqnarray}
where $p_1 ( \theta_1), p_2 ( \theta_2),q_1 (\theta_1),q_2 (\theta_2),r_1 
(\theta_1), r_2 (\theta_2)   $ are $\mathcal{N} =1$ superfields
in there own right. 
So, we can write the Lagrangian as 
\begin{equation}
  \mathcal{L} = D^2_1  [r_1 (\theta_1)]_{\theta_1 =0}= D^2_2 [r_2 (\theta_2)]_{\theta_2 =0}.
\end{equation}
The supersymmetry of this theory will be generated by the super-charges 
$
Q_{1a} = \partial_{1a}  - (\gamma^\mu \theta_1)_a \partial_\mu,$ and $  
Q_{2a} = \partial_{2a}  - (\gamma^\mu \theta_2)_a \partial_\mu
$. 
In absence of a boundary this theory is invariant  under the supersymmetry generated by both 
$Q_{1a} $ and $Q_{2a} $. However, in the presence of a boundary the supersymmetric transformations generated by both 
$Q_{1a}$ and $Q_{2a}$ generate boundary terms.  
So, on the boundary we can again   preserve only half of the total supersymmetry. Thus, with a boundary we can only preserve 
the supersymmetry generated by either
$\epsilon^{1+}Q_-$ or $\epsilon^{1-}Q_+$,  and by 
either $\epsilon^{2+}Q_-$ or $\epsilon^{2-}Q_+$. Now, after adding suitable boundary terms that preserve half of 
the supersymmetry, we get the following four possible Lagrangians 
\begin{eqnarray}
 \mathcal{L}^{1\pm 2 \pm} &=& d^{1\pm}d^{2\pm} 
[\Phi]_{\theta_1 = \theta_2 =0}, 
\end{eqnarray}
where 
\begin{eqnarray}
d^{1\pm}=(D^2_1 \pm \partial_3), && d^{2\pm}=(D^2_2 \pm \partial_3). \label{1} 
\end{eqnarray}
Now the  Lagrangian corresponding to $d^{1\pm} d^{2\pm}$ preserves the 
supersymmetry generated by  $\epsilon^{1\pm} Q_{1\mp}$ and $\epsilon^{2\pm} Q_{2\mp}$. 
It may be noted this Lagrangian  preserves only half the supersymmetry because the supersymmetry 
generated by   $\epsilon^{1\mp} Q_{1\pm}$ and $\epsilon^{2\mp} Q_{2\pm}$ is broken by it. 

\section{Matter-Born-Infeld Action}

In this paper we will consider the specific example of a matter-Born-Infeld
Lagrangian in $\mathcal{N} =2$ superspace, motivated by the potential
application to D2-branes ending on D4-branes in the presence of $RR$ fields.
This  Lagrangian   will be used for  analysing  the partial breaking of supersymmetry due to a 
combination of 
non-anticommutative deformations and boundary effects.
We first define two spinor superfields $\Gamma_{1a}$ and $\Gamma_{2a}$ which we
use to construct  covariant derivatives for matter fields $\Phi$ and $\bar \Phi$,
\begin{eqnarray}
 \nabla_{1a} \Phi =D_{1a}\Phi -i \Gamma_{1a} \Phi, &&
 \nabla_{2a} \Phi = D_{2a}\Phi -i \Gamma_{2a} \Phi,  \nonumber \\ 
 \nabla_{1a} \bar \Phi =D_{1a}\bar \Phi +i \bar \Phi\Gamma_{1a} , &&
 \nabla_{2a} \bar  \Phi = D_{2a}\bar \Phi +i \bar \Phi\Gamma_{2a}.
\end{eqnarray}
We can also construct the following field strengths from these spinor superfields
\begin{eqnarray}
 \omega_{1a} &=&  \frac{1}{2} D^b_1 D_{1a} \Gamma_{1b} - \frac{i}{2}  \{\Gamma^b_1, D_{1b} \Gamma_{1a}\}
- \frac{1}{6} [ \Gamma^b_1 ,
\{ \Gamma_{1b}, \Gamma_{1a}\}],\nonumber \\
 \omega_{2a} &=&  \frac{1}{2} D^b_2 D_{2a} \Gamma_{2b} - \frac{i}{2}  \{\Gamma^b_2, D_{2b} \Gamma_{2a}\}
- \frac{1}{6} [ \Gamma^2_1 ,
\{ \Gamma_{2b}, \Gamma_{2a}\}]. 
\end{eqnarray}
The Born-Infeld  Lagrangian can now be written as  \cite{1b, 2b, 3b, 4b}
\begin{eqnarray}
 \mathcal{L}_{bi} &=& D^2_1 [ \omega^a_1\omega_{1a}]_{\theta_1 =0} + D^2_2 
[\omega^a_2\omega_{2a}]_{\theta_2 =0} \nonumber \\ && 
+  D_1^2 D_2^2 [\omega^a_1\omega_{1a}\omega^b_2\omega_{2b} B (K_1, K_2)  ]_{\theta_1 = \theta_2 =0},
\end{eqnarray}
where $K_1 = D_1^2[\omega^a_1\omega_{1a}] $, $  K_2 = D_2^2 [\omega^a_2\omega_{2a}]$ and $B$ must satisfy a constraint equation.
For the Abelian Born-Infeld Lagrangian the constraint can be solved and
$B (K_1, K_2)$ can be written as 
\begin{eqnarray}
  B (K_1, K_2) = \frac{1}{2}\left[1- (K_1+ K_2)
 + \sqrt{4(1 - (K_1+ K_2 ) + (K_1 -K_2)^2 }\right]^{-1}.
\end{eqnarray}
Now to write the matter  Lagrangian   we define 
$
 \theta_a = ( \theta_{1a} -  i \theta_{2a}), \, 
 \bar \theta_a =  ( \theta_{1a} +  i \theta_{2a})
$, and
 $
 \partial_a =  ( \partial_{1a} +  i \partial_{2a})/2, \, 
 \bar \partial_a = ( \partial_{1a} -  i \partial_{2a})/2
$.
We similarly define 
$
D_a = \frac{1}{2} ( D_{1a} +  i D_{2a}), \, 
 \bar D_a = \frac{1}{2} ( D_{1a} -  i D_{2a})
$, and the covariant derivatives 
$
\nabla_a =  ( \nabla_{1a} +  i \nabla_{2a})/2, \, 
 \bar \nabla_a = ( \nabla_{1a} -  i \nabla_{2a})/2$. 
So, we can write  the  matter-Born-Infeld Lagrangian on a manifold without boundaries as
\begin{eqnarray}
 \mathcal{L} &=& D^2_1 D^2_2 [ \nabla^a \Phi  \bar \nabla_a  \bar  \Phi
 + \mathcal{V} [\Phi, \bar \Phi]]_{\theta_1 = \theta_2 =0}\nonumber \\&&
+ D^2_1 [ \omega^a_1\omega_{1a}]_{\theta_1 =0} + D^2_2 
[\omega^a_2\omega_{2a}]_{\theta_2 =0} \nonumber \\&&
+ D_1^2 D_2^2 [\omega^a_1\omega_{1a}\omega^b_2\omega_{2b} B (K_1, K_2)  ]_{\theta_1 = \theta_2 =0},
\end{eqnarray}
where $\mathcal{V} [\Phi, \bar \Phi]$ is a potential term.
This Lagrangian is invariant under the 
following gauge transformation,
\begin{eqnarray}
\Gamma_{1a} &\to& u\nabla_{1a} u^{-1}, \nonumber \\ 
 \Gamma_{2a} &\to& u\nabla_{2a} u^{-1}.\label{gauge}
\end{eqnarray} 
Now, in the presence of a boundary, we can preserve $\mathcal{N} = 1$
supersymmetry by modifying this Lagrangian to 
\begin{eqnarray}
 \mathcal{L}^{1\pm 2\pm}  &=& d^{1\pm}d^{2\pm} [ \nabla^a \Phi  \bar \nabla_a  \bar  \Phi
 + \mathcal{V} [\Phi, \bar \Phi]]_{\theta_1 = \theta_2=0} \nonumber \\&&
+ d^{1\pm} [\omega^a_1\omega_{1a}]_{\theta_1 =0} + d^{2\pm} [ \omega^a_2\omega_{2a}]_{\theta_2 =0}
\nonumber \\&&+  d^{1\pm}d^{2\pm} [ \omega^a_1\omega_{1a}\omega^b_2\omega_{2b} B (K_1, K_2)]_{\theta_1 = \theta_2=0}
, \label{a}
\end{eqnarray}
where 
$d^{1\pm}d^{2\pm} $ are given by Eq. (\ref{1}).
These Lagrangians are still invariant under the gauge transformation given by 
Eq. (\ref{gauge}). It may be noted that if the gauge part included Chern-Simons terms then 
this theory would not be gauge invariant,
but gauge invariance could be restored by the addition of further
boundary terms which would cancel the boundary piece generated by the gauge
transformation.
This has been considered in the context
of the ABJM model in \cite{fb, Chu:2009ms, gb} in component form, in
$\mathcal{N} = 1$ superspace, and for the Abelian case in $\mathcal{N} =2$
superspace. However, the full boundary action for the non-Abelian $\mathcal{N} = 2$
case has not yet been constructed.

\section{Boundary Supercharges and Boundary Superfields}
In this section we describe the relation between the bulk and boundary
supersymmetry \cite{eb, fb}.
For $\mathcal{N} =1$ supersymmetry, 
the bulk supercharge $Q_a$ can also be decomposed as 
$
 \epsilon^a Q_a = \epsilon (P^+ + P^-) Q \, 
= \epsilon^+ Q_- + \epsilon^- Q_+
$.
 These bulk supercharges can be written as 
$
 Q_- = Q'_- + \theta_- \partial_3, $ and $ 
 Q_+ = Q'_+ - \theta_+\partial_3
$.
where $Q'_{\pm}$ are the boundary 
supercharges given by  
$
 Q'_{\pm} = \partial_{\pm} - \gamma^s \theta_{\mp} \partial_s.
$
Here $s$ is the index for the 
coordinates along the boundary, i.e.\ compared to $\mu$ the case $\mu = 3$ is
excluded for a boundary at fixed $x^3$.
Now by definition $Q_{\pm}$ are the generators of the 
 half supersymmetry of the bulk fields and 
$Q'_{\pm}$ are the standard supersymmetry generators for the boundary fields. 
We also define $M_+ = \exp ( +  \theta_- \theta_+ \partial_3)$ and 
$M_-= \exp ( - \theta_+ \theta_- \partial_3)$ and let $M_+^{-1}$ and $M_-^{-1}$ be their inverses. Now we have
\begin{eqnarray}
 Q'_{-} &=& M_-^{-1}
 Q_- M_-, 
\nonumber \\
Q'_+ &=&M_+^{-1}
 Q_+M_+.
\end{eqnarray}
If we write  
\begin{equation}
 \phi = M_+ \phi'_+ \;\; \rm{or} \;\;
\phi = M_- \phi'_-
\end{equation}
where $\phi'_\pm$ are given in terms of boundary superfields $a'$ and $b'$ by 
$
 \phi'_+  = [ a' (\theta_-) + \theta_+ b' (\theta_-) ]$ or $ 
\phi'_- = [ a' (\theta_+) + \theta_- b' (\theta_+) ]
$, then
\begin{eqnarray}
 \epsilon^+ Q_- \phi  
= M_- {\epsilon^+}' Q'_- \phi'_- \;\; \rm{or} \nonumber \\ 
 \epsilon^- Q_+ \phi 
= M_+ {\epsilon^-}' Q'_+ \phi'_+,
\end{eqnarray}
where $Q'_- \phi'_- =  Q'_- a' (\theta_+) 
- \theta_- Q'_- b' (\theta_+)$ and $Q'_+ \phi'_+ = Q'_
+ a '(\theta_-) 
- \theta_+ Q'_+ b' (\theta_-) $. 
This gives the decomposition of $\phi$ into boundary superfields depending on
which supersymmetry is preserved.

Now, for $\mathcal{N} =2$ supersymmetry, 
the bulk supercharges  $Q_{na}$ (where $n =1, 2$) can also be decomposed as 
$
 \epsilon^{na} Q_{na} 
= \epsilon^{n+} Q_{n-} + \epsilon^{n-} Q_{n+}$, and
written as 
$
 Q_{n-} = Q'_{n- }+ \theta_{n-} \partial_3 $ and $ 
 Q_{n+} = Q'_{n+} - \theta_{n+}\partial_3, 
$
where $Q'_{n\pm}$  are the boundary 
supercharges given by  
$
 Q'_{n\pm} = \partial_{n\pm} - \gamma^s \theta_{n\mp} \partial_s
$. 
We again define 
$M_{n+} = \exp ( +  \theta_{n-} \theta_{n+} \partial_3)$ and 
$M_{n-}= \exp ( - \theta_+ \theta_- \partial_3)$ and let $M_{n+}^{-1}$ and $M_{n-}^{-1}$ be there inverses. Then
\begin{eqnarray}
 Q'_{n-} &=& M_{n-}^{-1}
 Q_{n-}M_{n-}, 
\nonumber \\
 Q'_{n+} &=& M_{n+}^{-1}
 Q_{n+}M_{n+}.
\end{eqnarray}
As for $\mathcal{N} = 1$, we write, depending on the supersymmetry preserved,
\begin{equation}
 \Phi = M_{2\pm}M_{1\pm} \Phi'_{2\pm 1\pm},
\end{equation}
where $\Phi'_{2 \pm 1\pm }$ decompose into boundary superfields.
Now we have one of the following:
\begin{eqnarray}
  \epsilon^{1-} Q_{1+} \Phi &=& M_{2\pm}M_{1+} 
 {\epsilon^{1-}}' Q_{1+}' \Phi'_{2\pm 1+}, \nonumber \\ 
  \epsilon^{1+} Q_{1-} \Phi &=& M_{2\pm}M_{1-} {\epsilon^{1+}}' Q_{1-}' \Phi'_{2\pm 1-}, \nonumber \\
  \epsilon^{2-} Q_{2+} \Phi &=& M_{2+}M_{1\pm} 
 {\epsilon^{2-}}' Q_{2+}' \Phi'_{2+ 1\pm}, \nonumber \\ 
  \epsilon^{2+} Q_{2-} \Phi &=& M_{2-}M_{1\pm} 
 {\epsilon^{2+}}' Q_{2-}' \Phi'_{2- 1\pm},
\end{eqnarray}
describing  the (preserved) supersymmetry transformation of the boundary superfields.

We will now analyse the superalgebra for a bulk $\mathcal{N} =2$ supersymmetric theory
 in the presence of a boundary.
In the absence of a boundary 
\begin{eqnarray}
 \{Q_{na}, Q_{mb}\} = 2 \gamma_{ab}^{\mu}\partial_\mu \delta_{nm},
 &&  \{D_{na}, D_{mb}\} =- 2 \gamma_{ab}^{\mu}\partial_\mu \delta_{nm}, \nonumber \\
 \{Q_{na}, D_{mb}\} = 0. && 
\end{eqnarray}
Now, defining $D_{n \pm a} = (P_\pm)_{a}^{\;b} D_{nb}$, and similarly for $Q_{n\pm a}$, we can write  the full superalgebra in
a form adapted to the presence of a boundary as 
\begin{eqnarray}
 \{Q_{n+ a}, Q_{m+ b}\} = 2 (\gamma_{ab}^{s}P_+)\partial_s \delta_{nm},
 &&  \{D_{n+a}, D_{m+b}\} =- 2 (\gamma_{ab}^{s}P_+)\partial_s\delta_{nm}, \nonumber \\
 \{Q_{n- a}, Q_{m- b}\} = 2 (\gamma_{ab}^{s}P_-)\partial_s \delta_{nm},
 &&  \{D_{n-a}, D_{m-b}\} =- 2 (\gamma_{ab}^{s}P_-)\partial_s \delta_{nm}, \nonumber \\
 \{Q_{n+a}, Q_{m-b}\} = -2 (P_{-})_{ab}\partial_3 \delta_{nm},
 &&  \{D_{n+a}, D_{m-b}\} = 2 (P_-)_{ab}\partial_3 \delta_{nm}, \nonumber \\
 \{Q_{n\pm a}, D_{m\pm b}\} = 0. && 
\end{eqnarray}
Contracting $D_{n-a}D_{n+b} = (P_-)_{ab} (\partial_3 -D^2) $ and $ D_{n+a}D_{n-b} = -(P_-)_{ab} (\partial_3 + D^2) $ with 
$C^{ab}$ and using $(P_-)_a^a =1$, we
 can also write Eq. (\ref{1}) as 
\begin{eqnarray}
d^{1+}=D_{1+}D_{1-}, && d^{2+}=D_{2+}D_{2-}, \label{a1} \\
d^{1-}=D_{1-}D_{1+},&& d^{2-} =D_{2-}D_{2+}.\label{a4}
\end{eqnarray}
Thus, we can see how the  Lagrangian with the measure $d^{1\pm}d^{2\pm}$ preserves
 the right amount of supersymmetry on the boundary. 
This is because the Lagrangian corresponding to Eq. (\ref{a}) 
can be written as 
\begin{eqnarray}
 \mathcal{L}^{1\pm 2\pm}  &=& D_{2\pm}D_{2\mp}D_{1\pm}D_{1\mp}
 [ \nabla^a \Phi  \bar \nabla_a  \bar  \Phi
 + \mathcal{V} [\Phi, \bar \Phi] ]_{\theta_1 = \theta_2=0}
 \nonumber \\&& +
D_{2\pm}D_{2\mp}D_{1\pm}D_{1\mp}[\omega^a_1\omega_{1a}\omega^b_2\omega_{2b} B (K_1, K_2)]_{\theta_{1\mp} = \theta_{2\mp} =0}
 \nonumber \\&&
+D_{2\pm}D_{2\mp}[ \omega^a_1\omega_{1a} ]_{\theta_{2\mp} =0}
+D_{1\pm}D_{1\mp}[ \omega^b_2\omega_{2b}]_{\theta_{1\mp} =0}. 
\end{eqnarray}
This Lagrangian is again invariant under the gauge transformation given by Eq. (\ref{gauge}).
We can write it in terms of boundary superfields as 
\begin{eqnarray}
 \mathcal{L}^{1\pm 2\pm}  &=& 
- D_{2\pm}'D_{1\pm}' [ \Psi'_{1\mp 2\mp} ]_{\theta_{1\mp} = \theta_{2\mp} =0}
\nonumber \\&&
+D_{2\pm}'[ \Psi'_{ 2\mp} ]_{\theta_{2\mp} =0}
+D_{1\pm}'[ \Psi'_{1\mp } ]_{\theta_{1\mp} =0}, 
\end{eqnarray}
where 
\begin{eqnarray}
 \Psi'_{1\mp 2\mp} &=& D'_{2\mp}D'_{1\mp}
[{\nabla^a}' \Phi ' \bar \nabla_a'  \bar \Phi'
 + \mathcal{V} [\Phi', \bar \Phi'] ]_{\theta_{1\mp} = \theta_{2\mp} =0} \nonumber \\ && 
+ D_{2\mp}'D'_{1\mp}
[{\omega^a}'_1\omega_{1a}'{\omega^b}'_2\omega_{2b}' B' (K'_1, K'_2)]_{\theta_{1\mp} = \theta_{2\mp} =0},
\nonumber \\
\Psi'_{ 2\mp}&=& D_{2\mp}' [ {\omega^a}'_2\omega'_{2a}]_{\theta_{2\mp} =0},
\nonumber \\
\Psi'_{ 1\mp}&=& D_{1\mp}' [ {\omega^a}'_1\omega'_{1a}]_{\theta_{1\mp} =0}.
\end{eqnarray}
The boundary measure   only  contains $D'_{2\pm}D'_{1\pm}$. Thus, on the boundary only  
the
supersymmetry generated by  ${\epsilon^{1\pm}}' Q_{1\mp}'$ and ${\epsilon^{2\pm}}'
 Q_{2\mp}'$ is preserved. Furthermore, on the boundary ${\epsilon^{1\pm}}' Q_{1\mp}'$ and ${\epsilon^{2\pm}}'
 Q_{2\mp}'$ act as independent supercharges. So, we obtain a boundary theory with either  $(1,1)$ supersymmetry  
or $(2,0)$ supersymmetry.

\section{Non-Anticommutativity}
In this section we will consider the effect of imposing non-anticommutativity
between the Grassmann coordinates \cite{Ferrara:2000mm, Klemm:2001yu, e, Britto:2003aj, f, Britto:2003ey, Britto:2003kg, Lunin:2003bm, Ivanov:2003te}.
While more general deformations of the super-Poincar\'e algebra are possible, we only consider non-anticommutativity, i.e.\ we do not consider
$[x^{\mu}, x^{\nu}] \ne 0$ or $[x^{\mu}, \theta] \ne 0$.

We first promote ${ \theta}^{n\pm a}$ 
 to operators $\hat{ \theta}^{n \pm a} $ and impose the most general 
form of non-anticommutativity for an $\mathcal{N} =2$ supersymmetric theory in three dimensions, 
\begin{equation}
 \{\hat \theta^{ n\pm a}, \hat \theta^{m\pm  b}\} = C^{na \pm   mb \pm   },
\end{equation}
where $C^{na+ nb+ } = C^{na- nb- } =0$. 
It may be noted that if we had started from a theory with 
$\mathcal{N} =1$ supersymmetric in three dimensions, it would not be possible to 
partially break the supersymmetry. This is because in that case 
the only non-anticommutative deformation that could take place would be $
 \{\hat \theta^{ + a}, \hat \theta^{-  b}\} = C^{a+  b-   }$ which would break
 all the supersymmetry of the theory. Hence, it is not possible to obtain in this way
a theory with $\mathcal{N} =1/2$ supersymmetry in three dimensions. However, 
we will show in the next section  we can obtain a theory 
with $\mathcal{N} =1/2$ supersymmetry in three dimensions
by combining non-anticommutativity 
with boundary effects.

We can write the Fourier transformation of a scalar superfield on the undeformed superspace as 
\begin{equation}
{\Phi} ( {\theta}^{n\pm } ) =
 \int d^4\pi  
\exp (-{\pi^{n \pm a} {\theta}_{n \pm a } }) \Phi (\pi_{n\pm }),
\end{equation}
where $\exp (-\pi^{n \pm a}{\theta}_{n \pm a} )  = \exp
 - (\pi^{1 + a} {\theta}_{1 + a} + \pi^{1 - a} {\theta}_{1 - a}+
\pi^{2 + a} {\theta}_{2 + a} + \pi^{2 - a}  {\theta}_{2 - a}) $.
Now, we can  
 use Weyl
ordering and also express the Fourier transformation of 
a scalar superfield on the deformed superspace as 
\begin{equation} 
\hat{\Phi} ( \hat{\theta}^{n \pm } ) =
 \int d^4\pi  
\exp (-{\pi^{n \pm a} \hat{\theta}_{n \pm a } })\Phi (\pi_{n\pm }) .
\end{equation}
Here we have considered the most general form for non-anticommutativity and it  breaks all
the supersymmetry. Non-anticommutative deformations
which partially break the supersymmetry   can be obtained from this general case
 by setting some of the
$C^{na \pm   mb \pm   }$  to zero. 
 We can express the product of two fields  
on this deformed superspace as
\begin{eqnarray}
\hat{\Phi}_1(\hat{\theta}^{n\pm })  \hat{\Phi}_2 
 (\hat{\theta}^{n\pm }) &=&
 \int d^4 \pi  d^4 \tilde \pi
\exp ( -( \pi +\tilde \pi)^{n\pm a}\hat \theta_{n\pm a})\nonumber \\
&& \times \exp \left( \Delta \right) 
\hat{\Phi}_1(\pi^{n\pm })  \hat{\Phi}_2
 (\tilde \pi^{n\pm }),
\end{eqnarray}
where $\Delta =- C^{na \pm   mb \pm   }\pi_{n\pm a}\tilde \pi_{n\pm b}/2$.
This motivates the  definition of  the star product
  between ordinary functions. Thus,   
the non-anticommutativity replaces all the product of fields by star products as follows \cite{Ferrara:2000mm, Klemm:2001yu, e}
\begin{eqnarray}
 \Phi_1 (\theta^{n\pm }) \star \Phi_2 (\theta^{n\pm } ) 
&=& \Phi_1 (\theta^{n\pm }) \exp \left(\frac{1}{2}C^{na \pm   mb \pm   } \overleftarrow{\partial}_{n\pm a} \overrightarrow{\tilde \partial}_{n\pm b} \right) 
\nonumber \\
 & & \times \Phi_2 (\tilde \theta^{n\pm })_{\tilde \theta^{n\pm } 
= \theta^{n\pm } }
\end{eqnarray}
with $\overrightarrow{\partial}_a \theta^b = \delta_a^b$ while
$\theta^b \overleftarrow{\partial}_a = - \delta_a^b$ etc.

Now the non-anticommutative 
 bulk supercharges  $Q_{na}$ can again be written as 
$
 \epsilon^{na} Q_{na} 
= \epsilon^{n+} Q_{n-} + \epsilon^{n-} Q_{n+}$.
  We now  define 
$M_{n+\star} = \exp ( +  \theta_{n-} \theta_{n+} \partial_3)_\star$ and 
$M_{n-\star}= \exp ( - \theta_+ \theta_- \partial_3)_\star$ and let $M_{n+\star}^{-1}$ 
and $M_{n-\star}^{-1}$ be their inverses.
Here  all the products are understood as star-products. 
Now we can write the relation between the bulk and boundary supercharges in this deformed superspace as,
\begin{eqnarray}
 Q'_{n-} &=& M_{n-\star}^{-1}\star
 Q_{n-} M_{n-\star }, 
\nonumber \\ 
 Q'_{n+} &=& M_{n+\star}^{-1}\star
 Q_{n+} M_{n+\star}.
\end{eqnarray}
Thus, we can also write a relation between the bulk and boundary superfields
 in this deformed superspace as, 
\begin{equation}
 \Phi = M_{2\pm\star}\star M_{1\pm\star}\star \Phi'_{2\pm 1\pm},
\end{equation}
where $\Phi'_{2 \pm 1\pm }$ are boundary superfields.

We have obtained boundary projections of the non-anticommutative superfields. 
Now we define non-anticommutative field strengths as 
\begin{eqnarray}
 \omega_{ 1a\star} &=&  \frac{1}{2} D^b_1 D_{1a} \Gamma_{1b} - \frac{i}{2}  \{\Gamma^b_1, D_{1b} \Gamma_{1a}\}_\star
- \frac{1}{6} [ \Gamma^b_1 ,
\{ \Gamma_{1b}, \Gamma_{1a}\}_\star]_\star,\nonumber \\
 \omega_{2a\star} &=&  \frac{1}{2} D^b_2 D_{2a} \Gamma_{2b} - \frac{i}{2}  \{\Gamma^b_2, D_{2b} \Gamma_{2a}\}_\star
- \frac{1}{6} [ \Gamma^2_1 ,
\{ \Gamma_{2b}, \Gamma_{2a}\}_\star]_\star. 
\end{eqnarray}
The Born-Infeld  Lagrangian can now be written as 
\begin{eqnarray}
 \mathcal{L}_{bi} &=& D^2_1 [ \omega^a_{1\star}\star \omega_{1a\star}]_{\theta_1 =0} + D^2_2 
[ \omega^a_{2\star}\star \omega_{2a\star}]_{\theta_2 =0} \nonumber \\ && 
+  D^2_1 D^2_2[ \omega^a_{1\star}\star \omega_{1a\star}\star\omega^b_{2\star}\star 
\omega_{2b\star}\star  B _\star (K_{1\star}, K_{2\star})]_{\theta_1 = \theta_2 =0}
\end{eqnarray}
where $K_{1\star} = D_1^2[\omega^a_{1\star}\star \omega_{1a\star}]$ 
and $K_2 = D_2^2 [\omega^a_{2\star}\star\omega_{2a\star}]$. 
Now  the Lagrangian for this non-anticommutative theory will be obtained by 
replacing all products of fields in the Lagrangian given by Eq. (\ref{a})
  with star-products,
\begin{eqnarray}
 \mathcal{L} &=& d^{1\pm}d^{2\pm} [
\nabla^a \star \Phi \star  \bar \nabla_a \star \bar \Phi
 + \mathcal{V} [\Phi, \bar \Phi]_{\star}]_{\theta_1 = \theta_2 =0} 
\nonumber \\ &&  + d^{1\pm}d^{2\pm} [
\omega^a_{1\star}\star \omega_{1a\star}\star \omega^b_{2\star}\star 
\omega_{2b\star}\star  B _\star (K_{1\star}, K_{2\star})]_{\theta_1 = \theta_2 =0}
\nonumber \\&&
d^{1\pm} [ \omega^a_{1\star}\star \omega_{1a\star}]_{\theta_1 =0} 
+ d^{2\pm} [ \omega^{2a\star}\star\omega_{2a\star} ]_{\theta_2 =0} .
\end{eqnarray}
Here the non-anticommutative potential term $\mathcal{V}[\Phi, \bar\Phi]_\star$ 
 is again obtained 
by replacing  the product of superfields in the original potential term by 
star products.  
We can again write it in terms of boundary superfields as 
\begin{eqnarray}
 \mathcal{L}^{1\pm 2\pm}  &=& - D_{2\pm}'D_{1\pm}'
 [ \Psi'_{1\mp 2\mp \star} ]_{\theta_{1\mp} = \theta_{2\mp} =0}
\nonumber \\&&
+D_{2\pm}'[ \Psi'_{ 2\mp\star} ]_{\theta_{2\mp} =0}
+D_{1\pm}'[ \Psi'_{1\mp \star} ]_{\theta_{1\mp} =0}, 
\end{eqnarray}
where 
\begin{eqnarray}
 \Psi'_{1\mp 2\mp \star} &=&
D'_{2\mp} D'_{1\mp} [{\nabla^a}' \star \Phi' \star  \bar \nabla_a'  \star \bar  \Phi'
 + \mathcal{V} [\Phi', \bar \Phi']_{\star} \nonumber \\ &&
+ {{\omega^a}_{1\star}}'\star \omega'_{1a\star}\star{{\omega^b}_{2\star}}'\star 
\omega_{2b\star}'\star  B '_\star (K'_{1\star}, K'_{2\star})]_{\theta_{1\mp} = \theta_{2\mp} =0}, 
\nonumber \\
\Psi'_{ 2\mp\star}&=& D'_{2\mp} [{\omega^a_{2\star}}'\star \omega_{2a\star}' ]_{\theta_{2\mp} =0},
\nonumber \\
\Psi'_{ 1\mp\star}&=& D'_{1\mp} [{\omega^a_{1\star}}'\star \omega_{1a\star}']_{\theta_{1\mp} =0}.
\end{eqnarray}
This Lagrangian is invariant under the non-anticommutative gauge transformation given by 
\begin{eqnarray}
\Gamma_{1a} &\to& u \star \nabla_{1a} \star  u^{-1}, \nonumber \\ 
 \Gamma_{2a} &\to& u\star\nabla_{2a} \star u^{-1}.
\end{eqnarray} 
The boundary measure corresponding to $d^{1\pm}d^{2\pm}$ contains only  $
D'_{2\pm}D'_{1\pm}$ and so the boundary effects again break half the 
 supersymmetry. However, now the non-anticommutativity also partially breaks 
supersymmetry. By combining the boundary effects with non-anticommutativity, 
it is possible to obtain theories with  $\mathcal{N} =1$ supersymmetry or $\mathcal{N} =1/2$ supersymmetry in the bulk. In the next 
session we will analyse various combinations of these boundary effects with non-anticommutativity. 
                                                                                                                                          
\section{Partially Breaking Supersymmetry}
Various amount of supersymmetry can be broken by a combination of boundary effects
 with non-anticommutativity. The projection of the generators of bulk supersymmetry 
again reproduces the correct generators of the boundary supersymmetry. Thus, we have one of
\begin{eqnarray}
  \epsilon^{1-} Q_{1+} \Phi &=& M_{2\pm \star}\star M_{1+\star } \star
 {\epsilon^{1-}}' Q_{1+}' \Phi'_{2\pm 1+}, \nonumber \\ 
  \epsilon^{1+} Q_{1-} \Phi &=& M_{2\pm\star }\star M_{1-\star } 
\star {\epsilon^{1+}}' Q_{1-}' \Phi'_{2\pm 1-}, \nonumber \\
  \epsilon^{2-} Q_{2+} \Phi &=& M_{2+\star }\star M_{1\pm\star } \star
 {\epsilon^{2-}}' Q_{2+}' \Phi'_{2+ 1\pm}, \nonumber \\ 
  \epsilon^{2+} Q_{2-} \Phi &=& M_{2-\star }\star M_{1\pm\star } \star
 {\epsilon^{2+}}' Q_{2-}' \Phi'_{2- 1\pm},
\end{eqnarray}
with the combination of supersymmetry generators which are left unbroken depending both on the choice 
of the boundary projection and the non-anticommutative deformation.

If all the remaining components of $C^{na\pm mb\pm}$ are non-zero then all supersymmetry is broken. 
In fact, all supersymmetry will also be broken if the non-zero components of $C^{na\pm mb\pm}$
 break 
the supersymmetry that is preserved by the boundary action. For example, after
 introducing the boundary, if the measure 
is changed to $d^{1+}d^{2+}$ then the supersymmetry corresponding to $Q_{1-}$
 and $Q_{2-}$ is left unbroken,
but if the non-anticommutativity is then imposed in such a way that  $C^{1a- 2b -}$
 is non-zero then all the supersymmetry will be broken. 

It is also possible to impose non-anticommutativity in such a way that it breaks the 
same supersymmetry that would be broken by the 
boundary. In this case half the supersymmetry of the original theory survives. Thus, if 
 the measure changes to $d^{1+}d^{ 2+}$ and  $C^{1a+ 2b+}$ is non-zero, 
then the supersymmetry corresponding to $Q_{1-}$ and $Q_{2-}$ remains unbroken.
So, we get 
an $\mathcal{N} = 1$ theory in the 
bulk which corresponds to $\mathcal{N} = (0,2)$ supersymmetry on the boundary,
with the star product defined with only $C^{1a+ 2b +}$ being non-zero.
Similarly, if 
 the measure changes to $d^{1-}d^{ 2-}$ and $C^{1a- 2b-}$ is non-zero, 
we preserve $\mathcal{N} = (2,0)$ supersymmetry on the boundary.
However, if the measure changes to $d^{1+}d^{ 2-}$ and  $C^{1a1+ 2b -}$ is non-zero, 
or if the measure changes to $d^{1-}d^{ 2+}$ and  $C^{1a- 2b +}$ is non-zero, 
 then
in both these cases we get
$\mathcal{N} = (1,1)$  
supersymmetry on the boundary.

The most interesting case is when half the supersymmetry
 left over 
after introducing the boundary is 
broken. For example, if the measure is changed to $d^{1+}d^{ 2+}$ and $C^{a1+ b 2-}$ 
is non-zero, only
the supersymmetry 
corresponding to $Q_{1-}$ is left unbroken.
 This corresponds to $\mathcal{N} = (0,1) $ on the
 boundary and thus 
$\mathcal{N} = 1/2$ in the bulk. Now for 
the same measure, if instead $C^{1a- 2b +}$ is non-zero,
 then the supersymmetry 
corresponding to $Q_{2-}$ is left unbroken which again corresponds to
  $\mathcal{N} = (0,1) $ 
on the boundary and
$\mathcal{N} = 1/2$ in the bulk.
 Similarly, if we change the measure to $d^{1-}d^{2-}$ and
 let either $C^{1a+ 2b -}$ or 
$C^{1a- 2b +}$
be non-zero, then $\mathcal{N} = (1, 0)$ supersymmetry is
 preserved on the boundary.
Further possibilities correspond to $d^{1-}d^{ 2+}$ or $d^{1+}d^{ 2-}$
with $C^{1a+ 2b +}$ or $C^{1a- 2b -}$ non-zero.                
Again the bulk theory
preserves $\mathcal{N} = 1/2$ supersymmetry  corresponding 
to $\mathcal{N} = (0,1) $
 or $\mathcal{N} = (1,0) $ on 
the boundary.

\section{Conclusion}
In this paper we have shown how a three-dimensional $\mathcal{N} = 1/2$
theory can be realized by starting with an $\mathcal{N} = 2$ theory and
breaking supersymmetry through a non-anticommutative deformation together with
the inclusion of a boundary. Most of the analysis is general but we also
discussed a Born-Infeld Lagrangian coupled to a matter field using
$\mathcal{N} =2$ superspace, motivated by the potential application to
D2-branes.

In summary, the supersymmetry of the $\mathcal{N} = 2$ theory was broken by  the presence of a boundary.
However, we modified the original theory by adding a boundary action to it such that the supersymmetric 
transformation of this boundary piece exactly cancels the boundary term generated from the supersymmetric transformation of the 
bulk theory. This way we were able to preserve half the supersymmetry of the original theory, i.e.\ $\mathcal{N} =1$
in three
dimensions. Depending on the choice of boundary action, this corresponds to a
two dimensional 
theory with  $\mathcal{N} = (1,1)$ or
$\mathcal{N} = (2,0)$ supersymmetry. 
We then analysed the
breaking of supersymmetry due to non-anticommutative deformations,
including the correct boundary projections of the bulk superfields in this general 
non-anticommutative superspace. We showed that, depending on the precise
anti-commutative deformation, it is possible to construct theories with 
$\mathcal{N} =1$ or
$\mathcal{N} =1/2$ supersymmetry  
in three dimensions. This was done by combining the breaking of supersymmetry by the  boundary  with the breaking of supersymmetry by the
non-anticommutativity. The reason both effects are required is that, unlike what happens in four dimensions, it is not possible to obtain a  
theory with $\mathcal{N} =1/2$ supersymmetry in three dimensions by only imposing non-anticommutativity. 

One interesting application of the methods in this paper would be to a
non-anticommutative deformation of a system of M2-branes with boundary on
an M5-brane.
 The low energy  Lagrangian for multiple
M2-branes is thought to be described by the ABJM Chern-Simons matter theory 
\cite{l1,l2,l3,l4}
which can be studied in $\mathcal{N} =2$ superspace.
The formalism developed in the present paper could be directly applied to
this system.
However, there is an additional complication that the
gauge transformation of the Chern-Simons Lagrangian generates a surface term.
 Thus, we would need to add another boundary 
term (or choose some suitable boundary conditions) such that the combined gauge variation of the original 
 theory along with this boundary piece will be gauge invariant. 
This has been done in the bosonic case \cite{Chu:2009ms} and in $\mathcal{N} =1$ superspace \cite{fb, gb, hb} where it
was shown that, not unexpectedly, the Lagrangian on 
the boundary is (a gauged version 
of) a WZW model.
It is 
expected that a  similar result will hold
in $\mathcal{N} =2$ superspace but the non-Abelian Chern-Simons action is
more complicated in this case, and the detailed construction of the
boundary action has not yet been
performed.

\end{document}